\documentclass[conference]{IEEEtran}
\usepackage{cite}
\usepackage{amsmath}
\usepackage{amssymb}
\usepackage[numbers]{natbib}
\usepackage{amsmath,amssymb,amsfonts}
\usepackage{algorithmic}
\usepackage{graphicx}
\usepackage{textcomp}
\usepackage{xcolor}
\usepackage{hyperref}

\usepackage{floatrow}
\usepackage{caption}

\usepackage{cuted}
\usepackage{flushend}

\usepackage[]{algorithm2e}
\usepackage{physics}

\begin{document}
\title{Constructing a NFT Price Index and Applications}

\author{
\IEEEauthorblockN{Hugo Schnoering}
\IEEEauthorblockA{\textit{Napoleon Group}\\
hugo.schnoering@napoleon-group.com}
\and
\IEEEauthorblockN{Hugo Inzirillo}
\IEEEauthorblockA{\textit{Napoleon Group}\\
\textit{CREST - Institut Polytechnique de Paris}\\
hugo.inzirillo@napoleon-group.com}
}

\maketitle

\begin{abstract}
    We are witnessing the emergence of a new digital art market, the art market 3.0. Blockchain technology has taken on a new sector which is still not well known, Non-Fungible tokens (NFT). In this paper we propose a new methodology to build a NFT Price Index that represents this new market on the whole. In addition, this index will allow us to have a look on the dynamics and performances of NFT markets, and to diagnose them.
\end{abstract}

\section{Introduction}
\begin{figure}[h!]
    \centering
    \includegraphics[width=0.9\columnwidth]{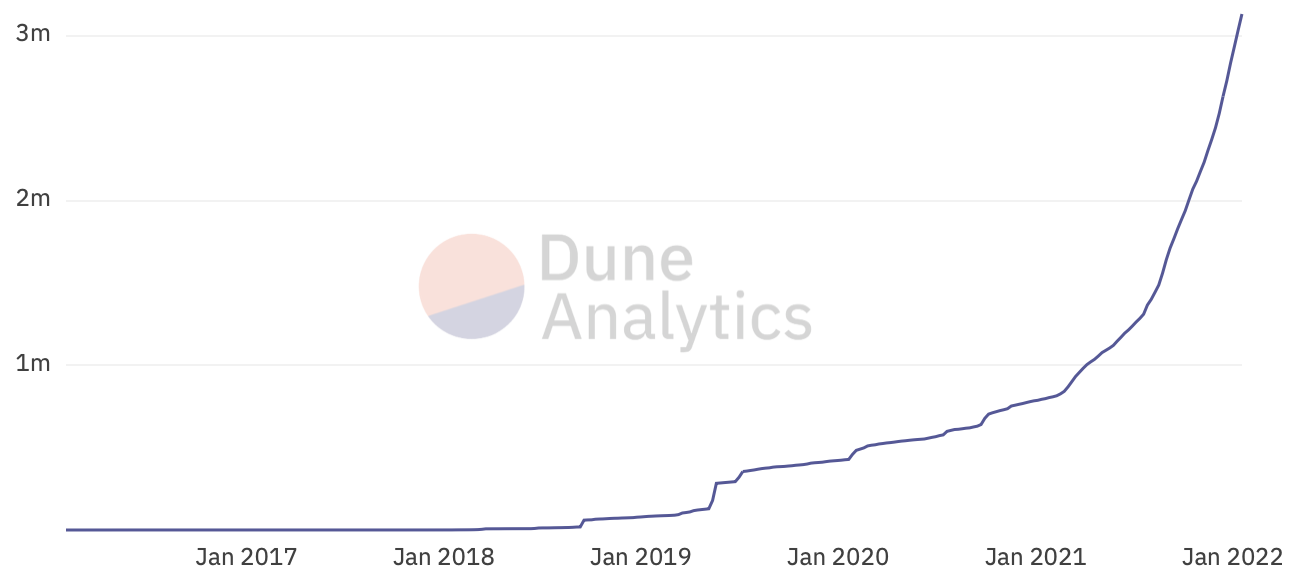}
    \caption{Cumulative number of wallet that have ever owned an ETH ERC-721 or ERC-1155 NFT \cite{thomasmdune}.}
    \label{fig:number_accounts}
\end{figure}

\begin{figure}[h!]
    \centering
    \includegraphics[width=0.9\columnwidth]{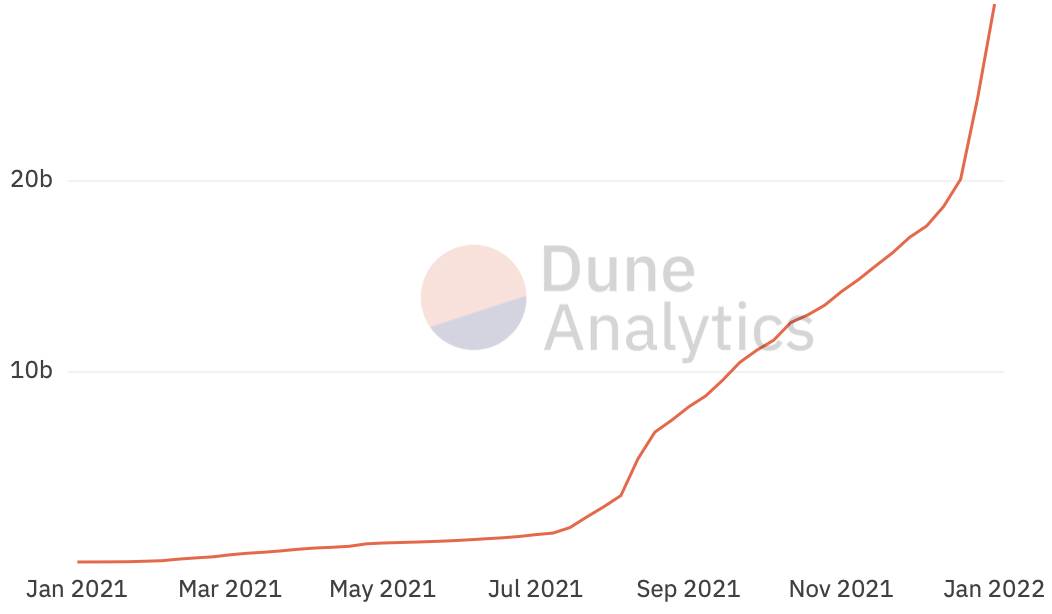}
    \caption{Cumulative number of NFT traded volumes in USD \cite{thomasmdune}.}
    \label{fig:volume_traded}
\end{figure}

Last year has been the year of the democratization of NFTs. The number of owners (figure \ref{fig:number_accounts}) and the number of transactions (figure \ref{fig:volume_traded}) on Ethereum have been exponentially increasing since January 2021. Moreover, the increasing interest of famous brand names for them also suggests that their adoption keeps growing. Online marketplaces like Opensea (\url{https://opensea.io}) or LooksRare (\url{https://looksrare.org}) have become focal points for trading these new assets. \\

NFTs or \textit{Non-Fungible Tokens} are transferable assets secured by a \textit{blockchain}. A {blockchain} is an ordered list of blocks that are linked together using cryptography. Blocks contain data about transactions, and are validated and added to the chain through a consensus protocol. Blockchain are used as public transaction ledgers of most cryptocurrencies. For instance bitcoin or ethereum respectively with the network Bitcoin \cite{nakamoto2008bitcoin} or Ethereum \cite{buterin2013ethereum}. The need for consensus of blockchain solves the double-spending problem using a cryptographic proof instead of a trusted central authority.  \\

Designed as a medium of exchange, cryptocurrencies are transferable assets as well. Because they are defined by their value, cryptocurrencies are perfectly fungible assets, for example a coin can be substituted to another. In opposition to cryptocurrencies, NFTs are uniquely identified by an id and a set of properties, and cannot be interchangeable or divisible. A NFT is thus the perfect way to represent anything unique on blockchains. They can be used for several purposes, but are often considered as digital art due to their properties. As art, most collectors acquire NFTs for aesthetic reasons, acquiring a social status or as a mean of investment \cite{kraussl2016there}.  \\

\begin{figure}[h!]
\RawFloats
\begin{minipage}{.49\linewidth}
    \centering
    \includegraphics[width=0.99\columnwidth]{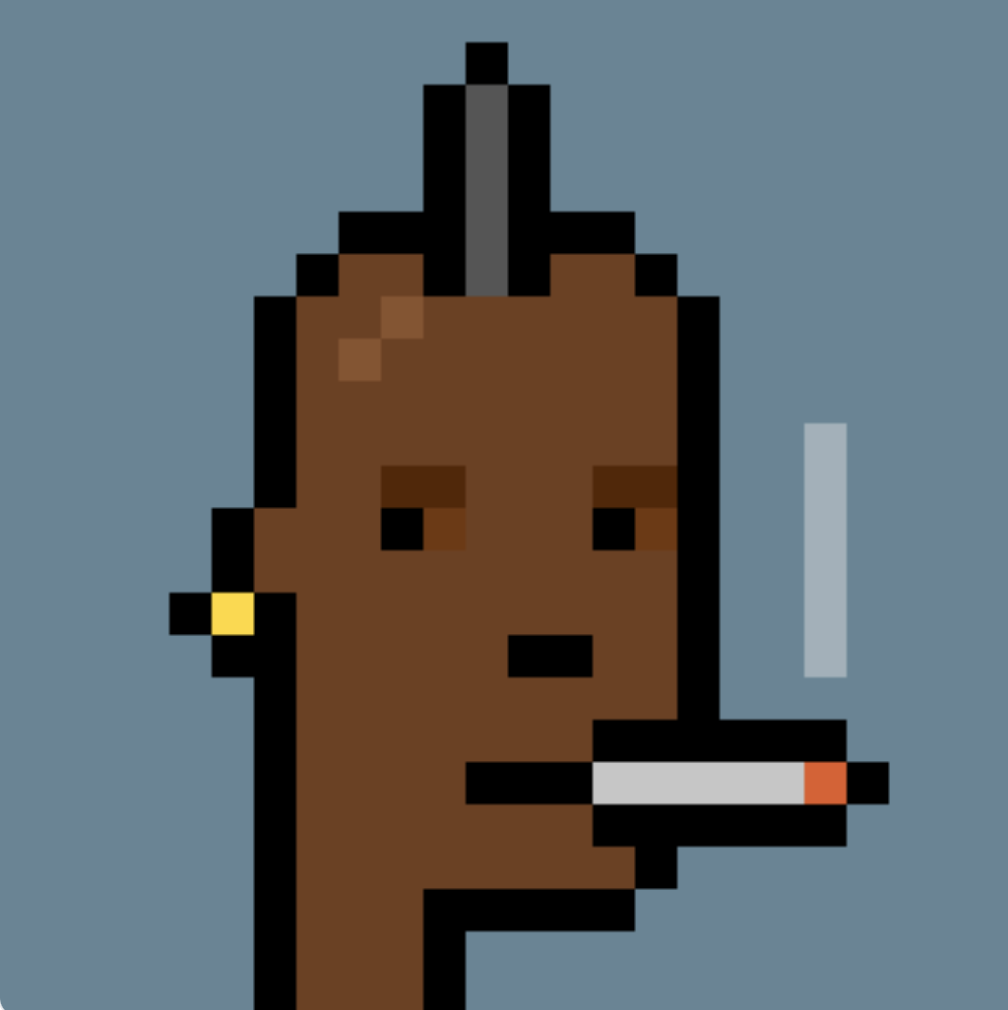}
    \caption{CryptoPunk \#1463. Properties and corresponding frequencies in the collection: male (60\%), cigarette (10\%), earring (25\%), Mohawk thin (4\%).}
    \label{fig:punk_1463}
    \end{minipage}%
    \hspace*{0.2cm}%
    \begin{minipage}{.49\linewidth}
    \centering
    \includegraphics[width=0.99\columnwidth]{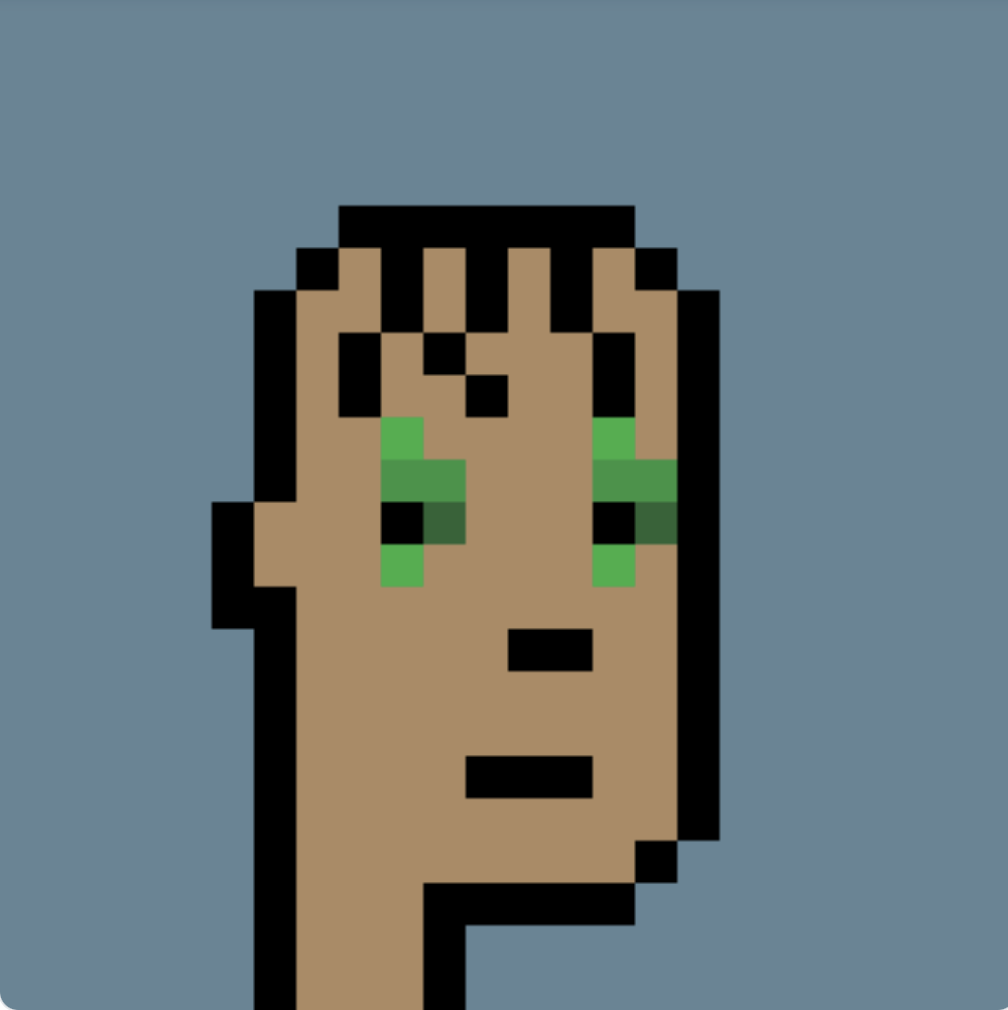}
    \caption{CryptoPunk \#1466. Properties and corresponding frequencies in the collection: male (60\%), stringy hair (5\%), clown eyes green (4\%).}
    \label{fig:punk_1466}
    \vspace*{0.4cm}
    \end{minipage}
\end{figure}

NFTs are created or \textit{minted} by the execution of a \textit{smart contract} setting in stone the creation of tokens. Several NFTs can be minted from the same contract, a collection is defined as a set of tokens minted from the same contract. Tokens minted from the same contract collect remain however perfectly identifiable and may differ from each others by their id and their properties or \textit{traits}. To illustrate this, we represent figures \ref{fig:punk_1463} and \ref{fig:punk_1466} respectively the tokens CryptoPunk \#1463 and CryptoPunk \#1466 of the Ethereum collection CryptoPunk. \\

Besides exceptional trade volumes, sales of NFTs have reached record prices, sometimes up to several tens of millions of dollars \cite{OpenseaCryptoPunk9998, OpenseaCryptoPunk3100}. Record breaking sales thus suggest a price explosion, but the dynamics of NFT markets are more complex. Indeed, if some top collections are steadily becoming more and more popular and expensive, e.g. the Ethereum CyberPunks and Bored Ape Yacht Club, prices of most collections vary widely. For this reason, it is difficult to have a global vision of the NFT market. In this work, we undertake the hedonic model to build a price index from thousands of NFT transaction records. Subsequent applications presented in this study include:
\begin{itemize}
    \item the detection of price explosion periods through statistical tests 
    \item the computation of the correlation matrix between the NFT returns with the return of the cryptocurrencies market 
    \item the detection of undervalued and overvalued assets
\end{itemize}

The hedonic regression framework is a well known model in Economics. Hedonic models decompose an item into its core characteristics and study the contribution of each of them to its value. These models are commonly used in Real Estate Economics \cite{herath2010hedonic, selim2011determinants}. A hedonic index is an index computed from a fitted hedonic regression model, \citeauthor{triplett2004handbook} reviews in \cite{triplett2004handbook} the different methodologies that can be used. Hedonic indices are often used as proxies of price indices, such as consumer price indices (CPI). A price index represents the aggregate price of a basket of items, and tracks how the prices of these items, taken as a whole, change over time. The hedonic methodology is particularly appraised because it can be applied to illiquid markets. Moreover, it can even handle the removal, replacement and addition of items over time. For all these reasons, several previous studies have applied it to diagnose art markets \cite{kraussl2008constructing, kraussl2016there, witkowska2014application}. Some works also used the hedonic model to price NFTs, but focused on a single collection: CryptoKitties \cite{kireyev2021infinite}, CryptoPunks \cite{kong2021alternative} or Decentraland \cite{goldberg2021economics}.  To the best of our knowledge, our article is the first proposal of a global NFT price index.

\section{Methods}

We will use the following notations:
\begin{itemize}
    \item $\mathcal{C}$ is the set of collections
    \item $\mathcal{A}$ is the set of assets
    \item $\{0, 1, ..., T\}$ is the set of sale dates
    \item $I(a, b) \triangleq \{a, a+1, ..., b-1, b\}$ 
    \item $I(b) \triangleq I(0, b)$
\end{itemize}

\subsection{Definitions}

An \textit{asset} $a$ is defined by a collection $C^{a}$, a token id $i^{a}$ and a dictionary of traits $\mathcal{P}^{a}$. $\mathcal{P}^{a}$ is a set of (trait, value) characterizing $a$ within its collection. $\mathcal{P}^{a}$ is usually used to quantify the scarcity of $a$ w.r.t. other assets of $C^a$, and, then, to price $a$. Intuitively the scarcer an asset is, the more expensive it should be. It is therefore straightforward to deal with frequencies of traits within a collection when it comes to build a pricing model.  Given an asset $a$, a couple $(p, v) \in \mathcal{P}^a$, the frequency of the value $v$ for the trait $p$ is defined as follows: 

\begin{equation}
    f^a_{(p, v)} \triangleq \frac{1}{|C^a|} \sum_{a^\prime \in C^{a}} \sum_{(p^\prime, v^\prime) \in \mathcal{P}^{a^\prime}} \mathrm{1}_{p^\prime=p} \mathrm{1}_{v^\prime=v}
\end{equation}

Since the number of traits may differ from an asset to another, even within the same collection, we construct three aggregate quantities: the minimum frequency $f^a_{\mathrm{min}}$ (equation \ref{eq:min_freq}), the average frequency $f^a_{\mathrm{avg}}$ (equation \ref{eq:avg_freq}), and the maximum frequency $f^a_{\mathrm{max}}$ (equation \ref{eq:max_freq}). All these quantities are set equal to 1 if $\mathcal{P}^{a}$ is empty.

\begin{equation}
f^{a}_{\min} \triangleq \min_{(p, v) \in \mathcal{T}^{a}} f^{a}_{(p, v)}
\label{eq:min_freq}
\end{equation}

\begin{equation}
f^{a}_{\mathrm{avg}} \triangleq \frac{1}{|\mathcal{P}^{a}|} \sum_{(p, v) \in \mathcal{P}^{a}} f^{a}_{(p, v)}
\label{eq:avg_freq}
\end{equation}

\begin{equation}
f^{a}_{\max} \triangleq \max_{(p, v) \in \mathcal{P}^{a}} f^{a}_{(p, v)}
\label{eq:max_freq} 
\end{equation}

A \textit{sale} is defined by an asset $a$, a date $t$ and a price $P^{a}_t$ in USD equivalent. 

\subsection{Pricing model}

The pricing model that we develop in this section is the result of three empiric observations:
\begin{itemize}
    \item as mentioned in the introduction, the price of an asset is impacted by the popularity and hype around its collection, 
    \item assets within the same collection may differ from each others by their traits, the scarcer the traits of an asset are, the more appreciated it is in its collection,
    \item the NFT market seems to experience periods during which prices of NFT are globally impacted positively or negatively. 
\end{itemize}

We finally come up with the multiplicative pricing model of equation (\ref{eq:pricing_model}).

\begin{equation}
    P^{a}_t = P \times f(C^{a}) \times g(a) \times h(t) \times \epsilon(a, t)
    \label{eq:pricing_model}
\end{equation}
where:
\begin{itemize}
    \item $P \in \mathbb{R}^{+*}$ defines a scale price,
    \item $f : \mathcal{C} \rightarrow \mathbb{R}^{+*}$ impacts the price of an asset according to its collection,
    \item $g : \mathcal{A} \rightarrow \mathbb{R}^{+*}$ impacts the price of an asset according to the scarcity of its traits within its collection,
    \item $h : \{0, 1, ..., T\} \rightarrow \mathbb{R}^{+*}$ impacts the prices according to the global state of the NFT market,
    \item $\epsilon : \{0, 1, ..., T\} \times \mathcal{A} \rightarrow \mathbb{R}^{+*}$ is a noise term explaining price fluctuations.
\end{itemize}
\vspace{0.3cm}
We model $f$, $g$, $h$ and $\epsilon$ as follows:
\begin{equation}
    f : C \rightarrow \exp  \sum_{C^{'}} \alpha_{C^{'}} \mathrm{1}_{C^{'} = C}
\end{equation}

\begin{equation}
    g : a \rightarrow \exp \left( \beta_{\min} f^{a}_{\min} + \beta_{\mathrm{avg}} f^{a}_{\mathrm{avg}} + \beta_{\max} f^{a}_{\max} \right)
\end{equation}

\begin{equation}
    h : t \rightarrow \exp \sum_{t^{'}=0}^T \gamma_{t^{'}} \mathrm{1}_{t^{'} = t}
\end{equation}

\begin{equation}
    \epsilon : (a, t) \rightarrow \exp \chi(a, t)
\end{equation}
where $\chi(a, t) \overset{i.i.d.}{\sim} \mathrm{Normal}(0, \sigma^2)$. \\

Equivalently, 

\begin{equation}
    \begin{split}
        \log P^{a}_t &= \log P + \sum_{C^{'}} \alpha_{C^{'}} \mathrm{1}_{C^{'} = C} \\
        &+ \beta_{\min} f^{a}_{\min} + \beta_{\mathrm{avg}} f^{a}_{\mathrm{avg}} + \beta_{\max} f^{a}_{\max} \\
        &+ \sum_{t^{'}=0}^T \gamma_{t^{'}} \mathrm{1}_{t^{'} = t} + \chi(t, a)
    \end{split}
\label{eq:hedonic}
\end{equation}

The model of equation (\ref{eq:hedonic}) is also known as the \textit{time dummy variable} version of the hedonic regression model \cite{triplett2004handbook}. Hedonic coefficients $P$, $\boldsymbol{\alpha}$ and $\boldsymbol{\beta}$ are inferred only once using all dates $\{0, 1, ..., T\}$, this approach is called the \textit{multi-period pooled regression} \cite{triplett2004handbook}. 

\subsection{Price Index Construction}

Since we used the time dummy variable method to train our pricing model, the hedonic index $I$ is defined as follows \cite{triplett2004handbook}: 

\begin{equation}
    \frac{I_{t+1}}{I_t} = \frac{\exp \gamma_{t+1}}{\exp \gamma_t}
    \label{eq:index}
\end{equation}

As a result, 

\begin{equation}
\small
    \begin{split}
        I_t &= A \prod_{t^{'}=0}^{t-1} \left( 1+ \left( \frac{I_{t^{\prime}+1}}{I_{t^{\prime}}}  -1 \right) \right) \\
        &= A \prod_{t^{'}=0}^{t-1} \left( 1+ \left(\frac{\exp \gamma_{t^{'}+1} }{\exp \gamma_{t^{'}} } -1 \right) \right) \\
        &= A \prod_{t^{'}=0}^{t-1} \frac{\exp \gamma_{t^{'}+1} }{\exp \gamma_{t^{'}} } \\
        &= A \exp ( \gamma_t - \gamma_0 )
    \end{split}
\end{equation}

Thus, the return of $I$ between $t$ and $t+1$ is $\frac{\exp \gamma_{t^{'}+1} }{\exp \gamma_{t^{'}} } -1$. \\

The construction of $I$ can be justified as follows, if $I$ is proportional to the geometric mean of prices over all assets of $\mathcal{A}$, then:

\begin{equation}
\small
    \begin{split}
        \frac{I_{t+1}}{I_t} &= \frac{\prod_{a \in \mathcal{A}} (P^a_{t+1})^{\frac{1}{|\mathcal{A}|}}}{\prod_{a \in \mathcal{A}}  (P^{a}_{t})^{\frac{1}{|\mathcal{A}|}}} \\
        &= \prod_{a \in \mathcal{A}} \left( \frac{P^a_{t+1}}{P^a_t} \right)^{\frac{1}{|\mathcal{A}|}} \\
        &= \frac{\exp \gamma_{t+1}}{\exp \gamma_t} \prod_{a \in \mathcal{A}} \exp \left(\chi(a, t+1) - \chi(a, t) \right)^{\frac{1}{|\mathcal{A}|}} \\
        &= \frac{\exp \gamma_{t+1}}{\exp \gamma_t} \exp \left( \frac{1}{|\mathcal{A}|} \sum_{a \in \mathcal{A}} \chi(a, t+1) - \chi(a, t) \right)
    \end{split}
\end{equation}

Since $\chi(a, t+1) - \chi(a, t) \overset{i.i.d.}{\sim} \mathrm{Normal}(0, 2 \sigma^2)$, according to the strong law of large numbers:

\begin{equation}
\small
    \frac{1}{|\mathcal{A}|} \sum_{a \in \mathcal{A}} \chi(a, t+1) - \chi(a, t) \underset{|\mathcal{A}| \rightarrow + \infty}{\rightarrow} 0, \ \mathrm{a.s.}
\end{equation}

Thus, 

\begin{equation}
\small
    \exp \left( \frac{1}{|\mathcal{A}|} \sum_{a \in \mathcal{A}} \chi(a, t+1) - \chi(a, t) \right) \underset{|\mathcal{A}| \rightarrow + \infty}{\rightarrow} 1, \ \mathrm{a.s.}
\end{equation}
because exp is continuous. Finally, if $\mathcal{A}$ is large enough:

\begin{equation}
\small
    \frac{I_{t+1}}{I_t} \approx \frac{\exp \gamma_{t+1}}{\exp \gamma_t}
\end{equation}

\subsection{Bubble Detection}
\label{sec:bubble_detection_methodo}

A bubble is often defined as a period of explosive or mildly explosive behavior in the price dynamic \cite{phillips2007limit}. In particular, autoregressive dynamics can be observed during speculative bubbles. For this reason, unit root tests are useful for making a market diagnosis, e.g. for detecting market excesses or mispricing. Augmented Dickey-Fuller (ADF) tests are usually used to determine whether a series $y$ is stationary or not, but they can also be used to detect explosive behaviors. Under the null hypothesis $H_0$, the process is autoregressive and has an unit root. Under the alternative hypothesis $H_1$, the process has an explosive root. Hypothesis $H_0$ is wide and must be specified in order to derive an asymptotic distribution and the critical values useful for the ADF test. To this end, \citeauthor{phillips2012testing} assume that the process $y$ is a random walk with an unit root and an asymptotically negligible drift \cite{phillips2012testing}, i.e.: 

\begin{equation}
\small
    y(t+1) = d(T+1)^{-\eta} + \theta y(t) + \epsilon(t)
    \label{eq:random_walk_drift}
\end{equation}
where $\theta=1$, $\eta > 1/2$ and $\epsilon(t) \overset{i.i.d.}{\sim} \mathrm{Normal}(0, \sigma^2)$. \\

The ADF testing procedure is applied to the regression model of equation (\ref{eq:reg_adf}).

\begin{equation}
\small
    \begin{split}
        \Delta y (t) = \mu  + \nu y(t-1) + \sum_{i=1}^k \psi^i \Delta y (t-i) + \epsilon(t)
    \end{split}
    \label{eq:reg_adf}
\end{equation}

where $\Delta y(t) = y(t) - y(t-1)$, $k \in \mathbb{N}$ is a lag order and $\epsilon \overset{i.i.d.}{\sim} \mathrm{Normal}(0, \sigma^2)$. The ADF statistic is defined as the t-statistic of the coefficient $\nu$ of the regression model. Under the null hypothesis (equation \ref{eq:random_walk_drift}), the process $y$ has an unit root, thus $\nu=0$. Under the alternative hypothesis of an explosive root, $\nu > 0$.   For this reason, we use the right-sided version of the ADF test, i.e. if ADF exceeds the critical value : $H_0$ is rejected and $H_1$ accepted, else : $H_0$ is accepted and $H_1$ rejected \\

In order to detect local bubbles and not to be fooled by pseudo-stationary behaviors, the ADF test can be performed on continuous sub-sample of $y$. We denote by $ADF_{t_1 \rightarrow t_2}$ the ADF statistic computed on $\{y_{t_1}, y_{t_1+1}, ..., y_{t_2} \}$. \citeauthor{phillips2012testing} develop in \cite{phillips2012testing} a methodology to detect multiple bubbles. For this purpose, they derive the BSADF statistic from the ADF statistic (equation \ref{eq:bsdaf}).
\begin{equation}
\small
BSDAF(t) = \sup_{t^{'} \in I(0, t-w)} ADF_{t^{'} \rightarrow t}
\label{eq:bsdaf}
\end{equation}
where $w$ is a hyper-parameter defining the minimum window size on which ADF tests can be  performed. The beginning and the end of a bubble, denoted respectively by $t_b$ and $t_e$, are detected as follows:

\begin{equation}
t_b \triangleq \inf \{t \in I(w, T), BSDAF(t) > v^{c}_{w}(t) \}
\end{equation}
\begin{equation}
t_e \triangleq \inf \{t \in I(t_b + \delta, T), BSDAF(t) < v^{c}_{w}(t) \}
\end{equation}

where $\delta$ is the minimum duration of a bubble and  $v^{c}_{w}(t)$ is a critical value defined as the $(1-c)$ quantile of the distribution of the statistic $\sup_{t^{'} \in \{w, ..., t\}} ADF_{0 \rightarrow t^{'}}$. Critical values $v^{c}_{w}(t)$ can be estimated using Monte-Carlo simulations. \\

\section{Data}
\begin{figure}[!h]
    \centering
    \includegraphics[width=0.95\columnwidth]{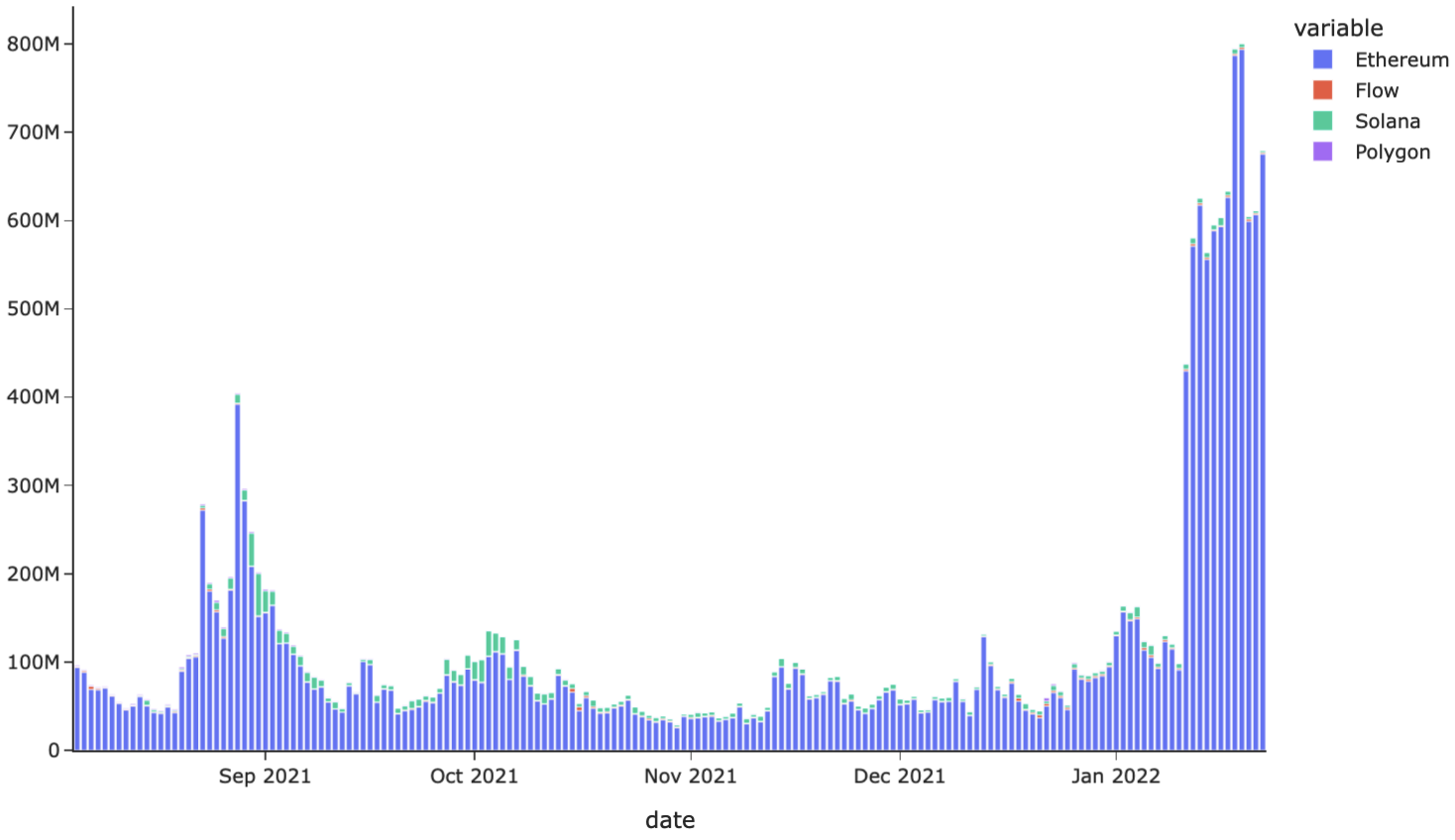}
    \caption{Daily sale volumes in dollar on four blockchains (data from \cite{cryptoslam}).}
    \label{fig:volume_chain}
\end{figure}

Several blockchains support NFTs including Ethereum, Solana, Flow, Tezos or Polygon. However, most of the NFT trade volume is concentrated on Ethereum (figure \ref{fig:volume_chain}). For this reason, we decide to focus our work on Ethereum NFT collections, more precisely on NFTs satisfying the standard ERC-721. \\

We have selected 59 Ethereum collections among the most traded collections on OpenSea during the second semester of 2021. The list of collection names and corresponding Ethereum smart contracts can be found in the appendix (section \ref{sec:smart_contracts}). \\

We use the official Opensea API (\url{https://docs.opensea.io}) to download the properties of all tokens in the selected collections, as well as all sale transactions concerning these collections from the 1st June 2021 to the 15th January 2022. Our dataset totals up 597198 transactions.

\section{Materials and Applications}

In order to train the model of equation (\ref{eq:reg_adf}) we use the Huber Regressor, a linear regression model that is robust to outliers. We train two models: a model with all collected collections in order to build the NFT price index, and another model with only metaverse-related collections to build the metaverse price index. Collections used for both indices can be found in the appendix (section \ref{sec:smart_contracts}). The first value of our indices, i.e. the hyper-parameter $A$, is set equal to 100. We report on table \ref{tab:asset_coef_hedonic} the inferred parameters $\boldsymbol{\beta}$ of $g$, the asset rarity impact function, for the NFT price index. \\

\begin{table}
    \centering
    \begin{tabular}{c|c}
    Coefficient & Value   \\
    \hline
    $\beta_{\min}$ & -0.176 \\
    $\beta_{\mathrm{avg}}$ & -0.009  \\
    $\beta_{\max}$ & -0.025 \\
    \end{tabular}
    \caption{Asset-related coefficients of the model. }
    \label{tab:asset_coef_hedonic}
\end{table}

\begin{figure*}[!ht]
\RawFloats
\begin{minipage}{.49\linewidth}
    \centering
    \includegraphics[width=0.99\columnwidth]{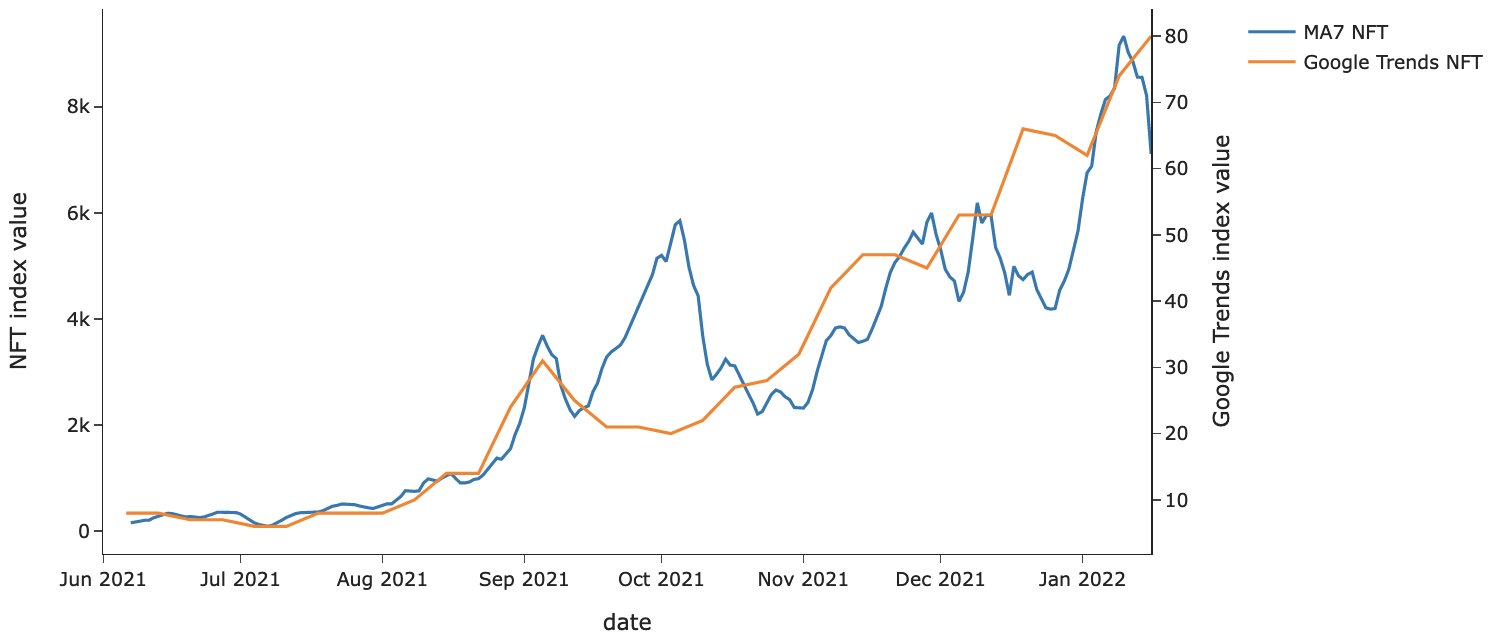}
    \caption{7-day-MA NFT Index.}
    \label{fig:nft_index}
    \end{minipage}%
    \begin{minipage}{.49\linewidth}
    \centering
    \includegraphics[width=0.99\columnwidth]{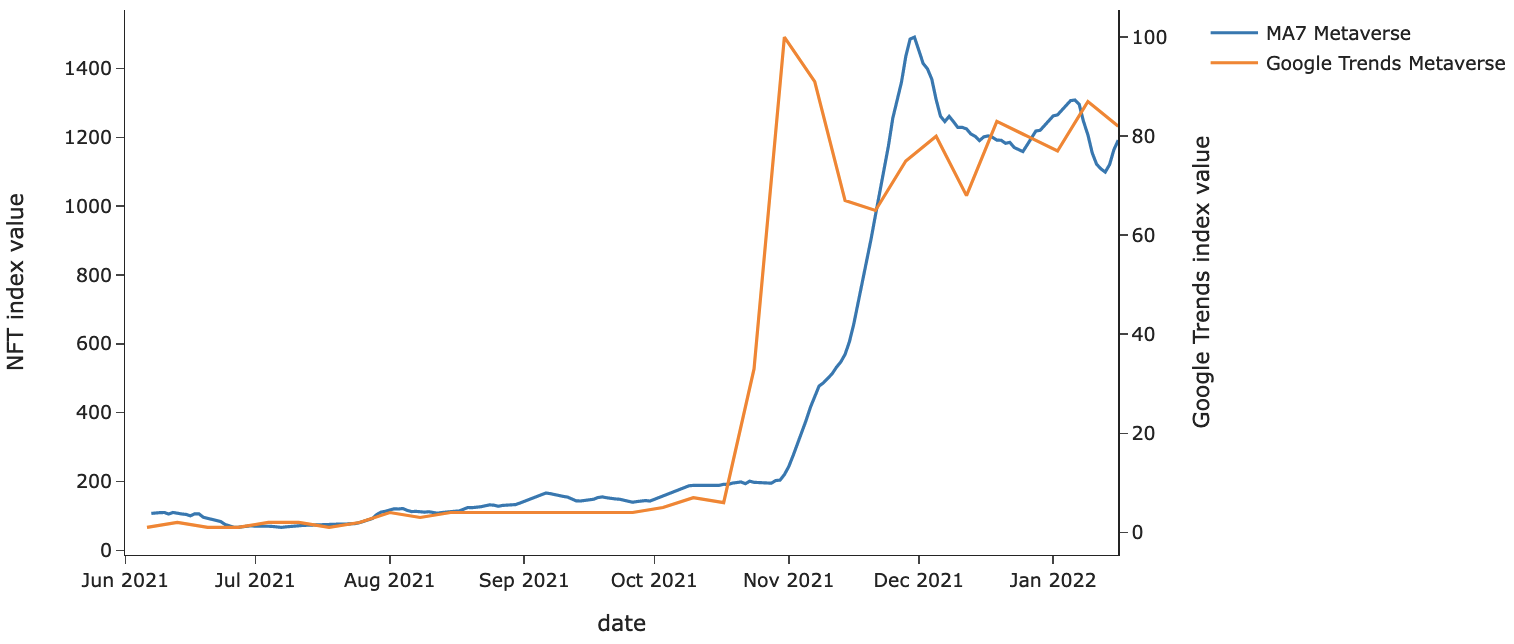}
    \caption{7-day-MA Metaverse Index. }
    \label{fig:metaverse_index}
    \end{minipage}
\end{figure*}

We plot in figures \ref{fig:nft_index} and \ref{fig:metaverse_index} the 7-day moving averages of the NFT and Metaverse Index respectively. Raw indices can be found in the appendix (section \ref{sec:raw_signals}). We also plot in figures \ref{fig:nft_index} and \ref{fig:metaverse_index} the Google Trends (GT) signals for the queries "nft" and "metaverse" between the 1st June 2021 and the 16th January 2022 \cite{googleTrendsMetaverse, googleTrendsNFT}. Since the GT signals for $q$ is proportional to the Google search volume for $q$, both GT signals will be used to chart the fad for NFTs and the metaverse. 

\subsection{Correlation with cryptocurrency markets}

In table \ref{tab:corr_matrix}, we report the correlation coefficients between the daily returns of our indices, ETH, BTC and SOL. \\

\begin{table}[!ht]
    \centering
    \begin{tabular}{c|ccccc}
         & NFT & Metaverse & ETH & SOL & BTC  \\
         \hline
         NFT & & 0.09 & 0.16 & 0.18 & 0.10 \\
         Metaverse & & & 0.30 & 0.17 & 0.20 \\
         ETH & & & & 0.59 & 0.82\\
         SOL & & & & & 0.48 \\
    \end{tabular}
    \caption{Correlation matrix of daily returns. Correlation coefficients are computed using the data from the 1st June 2021 to the 15th January 2022.}
    \label{tab:corr_matrix}
\end{table}

We report in table \ref{tab:return_vector} the realized return of our indices, ETH, BTC, and SOL between the 1st June 2021 and the 15st January 2022. 

\begin{table}
    \centering
    \begin{tabular}{c|c}
    & Return \\
    \hline
    NFT & 7011\%  \\
    Metaverse & 1064\% \\ 
    ETH & 26\% \\
    BTC & 6\% \\
    SOl & 372\% \\
    \end{tabular}
    \caption{Realized return computed from prices in USDT between the 1st June 2021 and the 15st January 2022.}
    \label{tab:return_vector}
\end{table}

\subsection{Bubble Detection}

\begin{figure*}[!ht]
\RawFloats
\begin{minipage}{.49\linewidth}
    \centering
    \includegraphics[width=0.95\columnwidth]{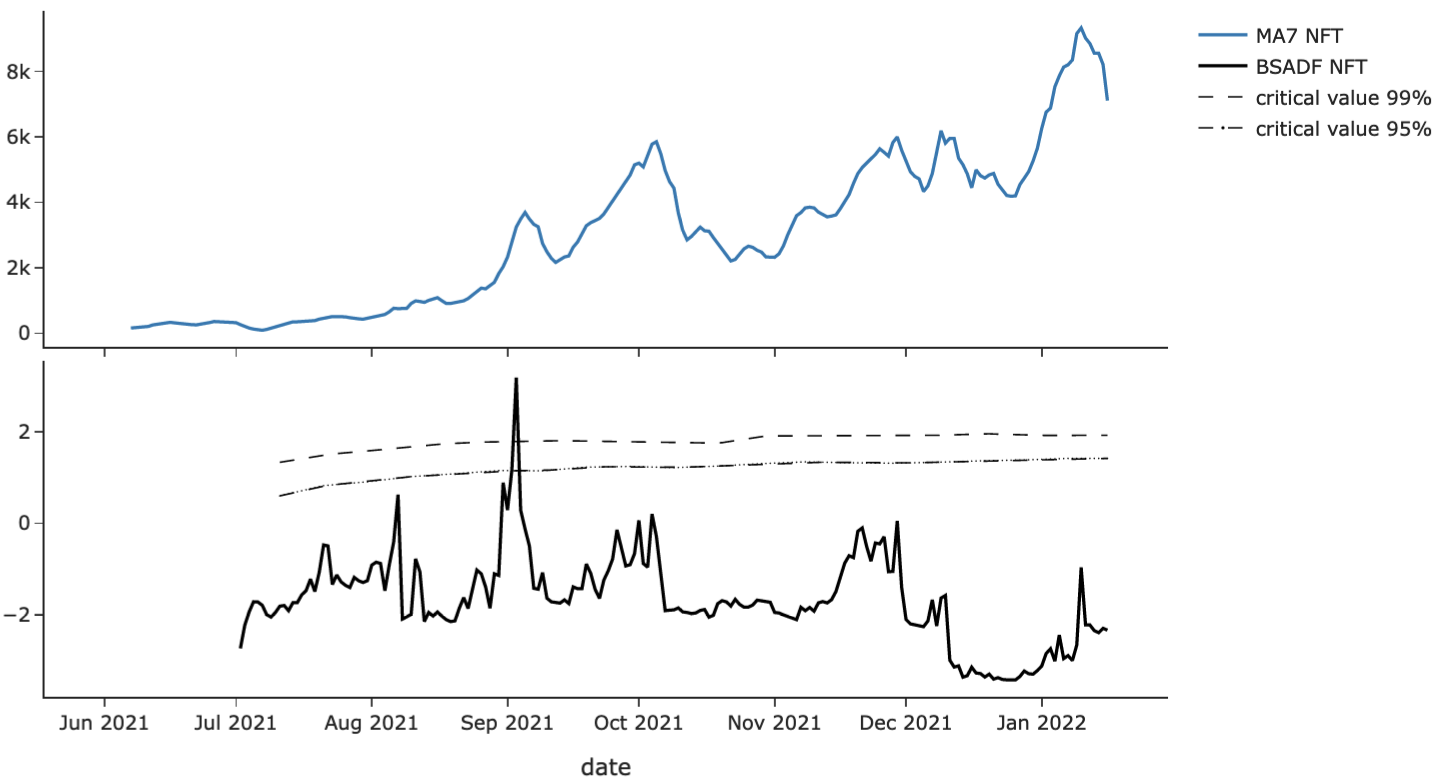}
    \caption{7-day-MA NFT Index (above). BSADF statistic signal for the NFT Index (below). $95\%$ and $99\%$ critical values (below).}
    \label{fig:bsadf_nft_index}
    \end{minipage}%
    \vspace{0.2cm}
    \begin{minipage}{.49\linewidth}
    \centering
    \includegraphics[width=0.95\columnwidth]{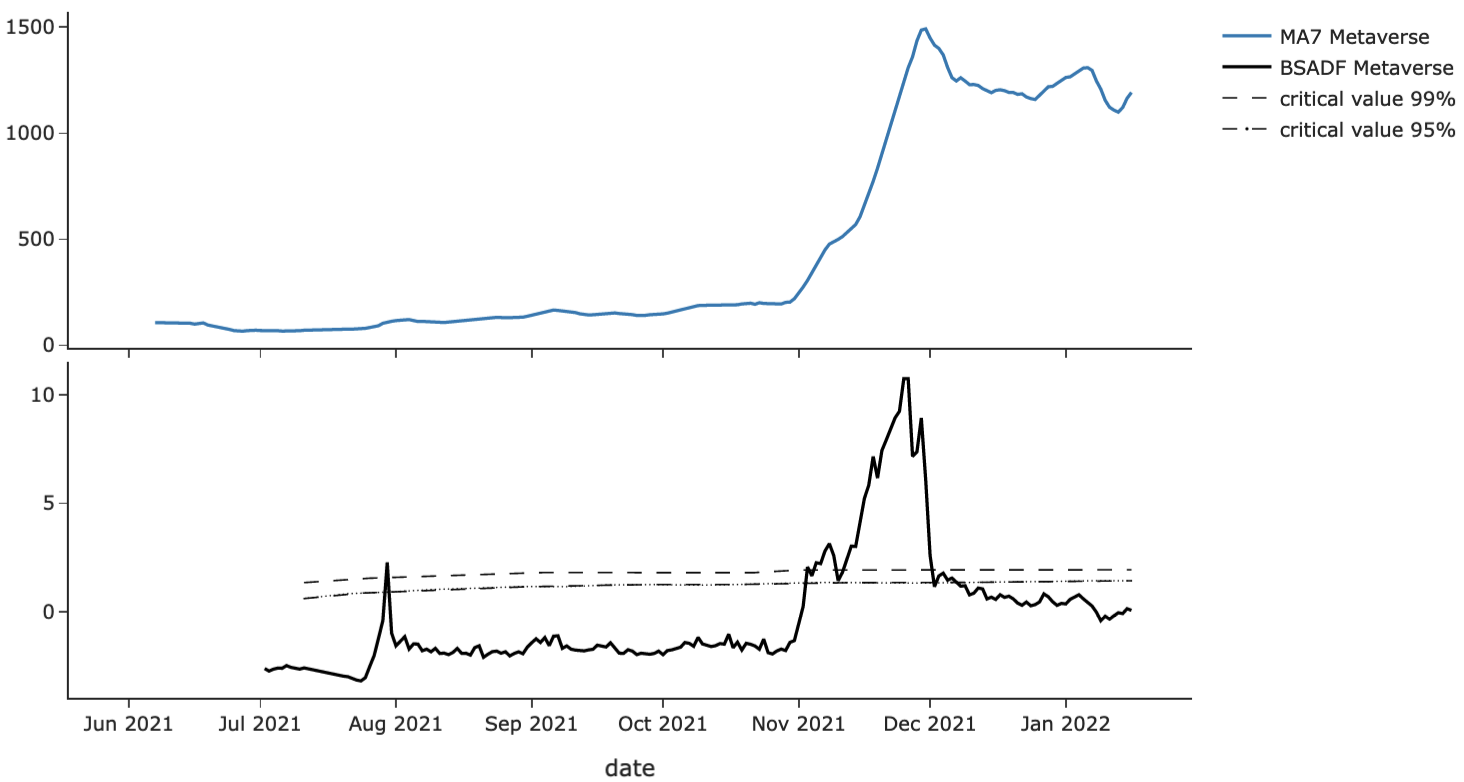}
    \caption{7-day-MA Metaverse Index (above). BSADF statistic signal for the Metaverse Index (below). $95\%$ and $99\%$ critical values (below).}
    \label{fig:bsadf_metaverse_index}
    \end{minipage}
\end{figure*}

The minimum length $w$ to compute an ADF test is fixed equal to $40$. We use the methodology of \citeauthor{phillips2012testing} presented in section \ref{sec:bubble_detection_methodo} to detect the start and the end of bubbles. $c$-critical values $v^{c}_{w}(t)$ of $\sup_{t^{'} \in I(w, t)} ADF_{0 \rightarrow t^{'}}$ are estimated using Monte-Carlo simulations of $N_{MC} = 5000$ random walks with an asymptotically negligible drift (equation \ref{eq:random_walk_drift}) where we set $\eta=\sigma=d=1$. $v^{c}_{w}(t)$ are estimated for $t \in I(w, T)$. We report in particular the $90\%$, $95\%$ and $99\%$ critical values when $t=T=230$ in table \ref{tab:critical_values}. We plot on figures \ref{fig:bsadf_nft_index} and \ref{fig:bsadf_metaverse_index} the $t \rightarrow BSADF(t)$ signals (equation (\ref{eq:bsdaf})) computed for both indices. We have also plotted the $95\%$ and $99\%$ critical value signal for the statistics $t \rightarrow \sup_{t^{'} \in I(w, t)} ADF_{0 \rightarrow t^{'}}$. We recall that an explosive behavior is detected in the price dynamic if the BSADF statistic exceeds the $99\%$ critical value signal. 

\begin{table}
    \centering
    \begin{tabular}{c|c}
    $c$ & $v^{c}_{w, t}$  \\
    \hline
    $90\%$ & 1.08 \\
    $95\%$ & 1.36 \\
    $99\%$ & 1.97 \\
    \end{tabular}
    \caption{Critical values $v^{c}_{w}(t)$ of SADF with $t=T=230$ and $w=40$.}
    \label{tab:critical_values}
\end{table}

\subsection{Undervalued and Overvalued Assets}

Once the parameters $P$, $\boldsymbol{\alpha}$, $\boldsymbol{\beta}$, $\boldsymbol{\gamma}$ and $\sigma$ of equation (\ref{eq:hedonic}) are inferred after a training phase, any asset $a$ can be priced at time $t \in \{0, ..., T\}$ if at least one asset of the same collection has been used during the training phase. According to the model of equation (\ref{eq:hedonic}), the log-price of $a$ at time $t$ follows the distribution $\mathrm{Normal}(\log\Bar{P}^a_t, \sigma^2)$ where $\log\Bar{P}^a_t$ is computed as follows:

\begin{equation}
\small
\begin{split}
\log \Bar{P}^{a}_t &= \log P +  \alpha_{C^{a}} +  \beta_{\min} f^{a}_{\min} + \beta_{\mathrm{avg}} f^{a}_{\mathrm{avg}} + \beta_{\max} f^{a}_{\max} + \gamma_{t} 
\end{split}
\label{eq:mean_log_price}
\end{equation}

We define the probabilities of being undersold and oversold in equations (\ref{eq:prob_undersold}) and (\ref{eq:prob_oversold}), respectively. 

\begin{equation}
    p_{\mathrm{under}}(P^{a}_t) \triangleq \frac{1}{2} \left( 1 - \mathrm{erf} \left( \frac{\log(P^{a}_t) - \log(\Bar{P}^{a}_t)}{\sigma \sqrt{2}} \right) \right)
\label{eq:prob_undersold}
\end{equation}

\begin{equation}
    p_{\mathrm{over}}(P^{a}_t) \triangleq \frac{1}{2} \left( 1 + \mathrm{erf} \left( \frac{\log(P^{a}_t) - \log(\Bar{P}^{a}_t)}{\sigma \sqrt{2}} \right) \right)
\label{eq:prob_oversold}
\end{equation}
where erf is the Gauss error function. 
According to the definitions (\ref{eq:prob_undersold}) and (\ref{eq:prob_oversold}), it is straightforward to derive the following properties:
\begin{itemize}
    \item $p_{\mathrm{over}} + p_{\mathrm{under}} = 1$,
    \item at $P^{a}_t=\Bar{P}^{a}_t$, $p_{\mathrm{over}}(P^{a}_t) = p_{\mathrm{under}}(P^{a}_t) = \frac{1}{2}$,
    \item $p_{\mathrm{under}}$ is strictly increasing, $\lim_{0^+} p_{\mathrm{under}} = 1$ and $\lim_{\infty} p_{\mathrm{under}} = 0$.
\end{itemize} 

The simplest investment strategy could consist in investing at time $t$ in an asset $a$ for which the listing price $P^a_t$ gives a high undersold probability $p_{\mathrm{under}}(P^a_t)$ below $1/2$, and listing it at $\Bar{P}^a_t$. One problem, however, is that equation (\ref{eq:mean_log_price}) assumes to have estimated $\gamma_t$, i.e. the market impact coefficient of today. To deal with it, $\gamma_t$ can be estimated using earlier in the day transactions, or it can be approximated by a moving average of the market impact coefficients over recent previous days.

\section{Discussion}
From figures \ref{fig:nft_index} and \ref{fig:metaverse_index}, we detect a strong upward trend for both price indices indicating a global price rise. This increase in prices appears to follow the rising interest for both themes. The NFT price index seems to have experienced several brief explosive periods, while the metaverse price index experienced an unique explosive period beginning in November 2021. This comes a short time after the renaming of Facebook's parent company to Meta reveiling their interest for the metaverse. \\

As shown in table \ref{tab:asset_coef_hedonic}, the parameters of $g$ are all non positive. As a consequence, $g$, and then the price, is strictly decreasing with the quantities $f_{\min}$, $f_{\mathrm{avg}}$ and $f_{\max}$. It statistically confirms our intuition that scarcer assets are more expensive. \\

According to table \ref{tab:corr_matrix}, returns of both indices are positively correlated with the returns of the top-tier cryptocurrencies BTC, ETH and SOL, but correlation coefficients remain relatively small compared to the ones between BTC, ETH and SOL. From table \ref{tab:return_vector}, we observe that the selected collections have, on the whole, dramatically outperformed these cryptocurrencies. Putting these results together suggests that blue ship NFTs have, taken as a whole, offered both-crypto-and-NFT investors high returns while diversifying their risk. \\

According to figure \ref{fig:bsadf_nft_index}, only a brief bubble has been detected by the presented methodology. Thus, the NFT price index has globally never experienced any explosive price dynamic, but it doesn't mean that some collections have not. Indeed, as shown in figure \ref{fig:bsadf_nft_index}, a bubble has been detected in the metaverse index between the 2. November 2021 and the 2. December 2021 following the renaming of Facebook. \\

Concerning our methodology, two weaknesses have been identified. First, the selection of collections is highly subjective and includes a look-ahead bias. Secondly, hedonic coefficients $\boldsymbol{\alpha}$ and $\boldsymbol{\beta}$ are hold fixed over time, thus, our model can not take into account the time varying popularity of collections to explain price dynamics. Both of these problems could be solved by implementing a quantitative rule to update the collection set on a regular basis, and, by using the \textit{adjacent dummy variable} method \cite{triplett2004handbook}, an alternative training procedure for the time dummy variable hedonic model. It could also be interesting in a future work to consider the temporal dependence of the trends according to the collections, or to detect NFT market regimes. Finally, it would also be appropriate to test and compare other bubble detection models. 

\section{Acknowledgement}

We would like to thank Opensea for providing us an API key to collect all the necessary data to conduct this study.

\section{Conclusion}

In this work, we have proposed a methodology to construct price indices for the NFT markets. In contrary to previous works, our indices aggregate transactions from varied collections, this allows us to better represent this new art market 3.0.  These indices have allowed us to analyse the dynamics and performances of NFT markets, and to perform diagnostic tests. In particular, we have used statistical tests to detect speculative bubbles. Finally, we have demonstrated that simple intuitive investment strategies could be derived from the pricing model used for the contruction of our indices.

\bibliographystyle{IEEEtranN}
\bibliography{bib}

\section*{Appendix}

\subsection{Raw signals}
\label{sec:raw_signals}

\begin{figure}[h!]
    \centering
    \includegraphics[width=0.99\columnwidth]{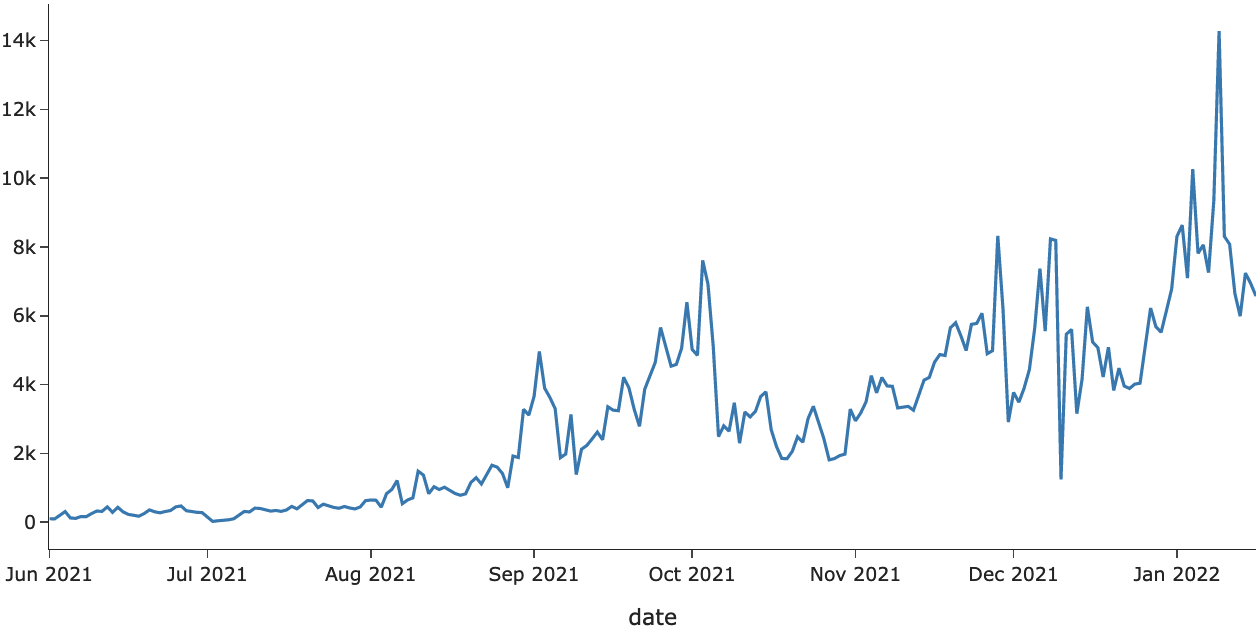}
    \caption{NFT Index.}
    \label{fig:raw_nft_index}
\end{figure}

\begin{figure}[h!]
    \centering
    \includegraphics[width=0.99\columnwidth]{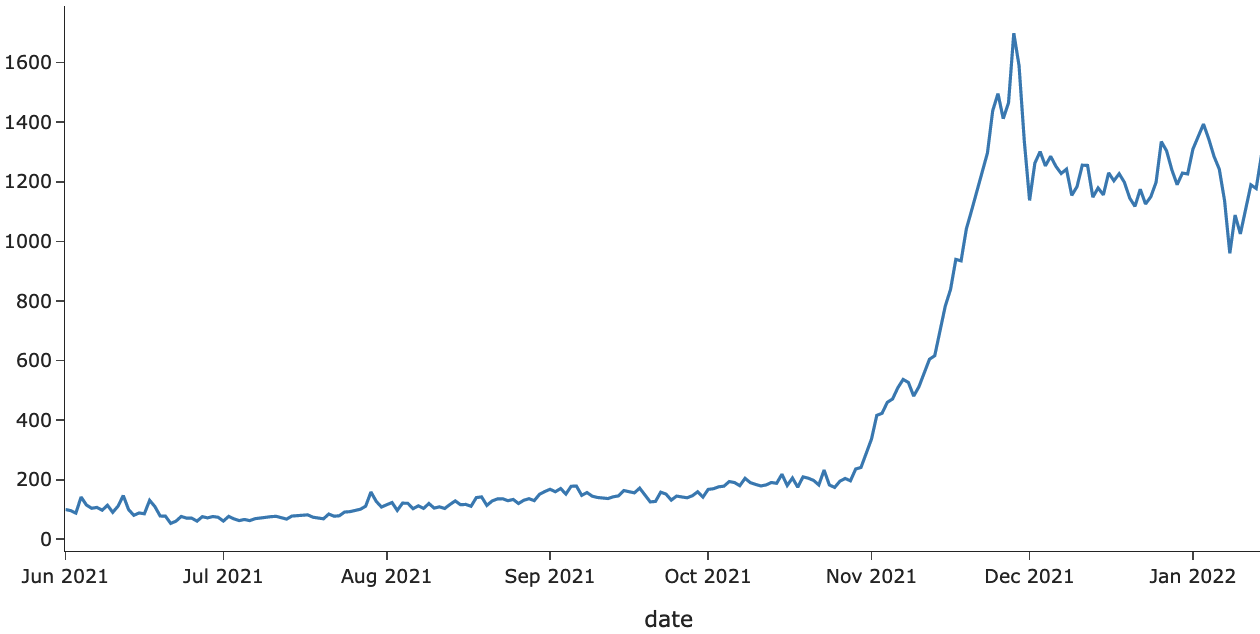}
    \caption{Metaverse Index.}
    \label{fig:raw_metaverse_index}
\end{figure}

\subsection{Smart contracts}
\label{sec:smart_contracts}

\begin{table}
    \tiny
    \centering
    \begin{tabular}{c|cc}
    Index & Collection & Ethereum Contract  \\
    \hline
    NFT
      & CryptoPunks & 0xb47e3cd837dDF8e4c57F05d70Ab865de6e193BBB \\
      & CloneX & 0x49cF6f5d44E70224e2E23fDcdd2C053F30aDA28B \\
      & MutantApeYachtClub &  0x60E4d786628Fea6478F785A6d7e704777c86a7c6 \\
      & BoredApeYachtClub &  0xBC4CA0EdA7647A8aB7C2061c2E118A18a936f13D \\
      & Sandbox & 0x50f5474724e0Ee42D9a4e711ccFB275809Fd6d4a \\
      & VOX Series 2 & 0xf76179bb0924BA7da8E7b7Fc2779495D7A7939d8 \\
      & Nanopass & 0xf54cC94f1F2f5De012B6Aa51F1E7eBdc43Ef5afC \\
      & Shiba Social Club &  0xD692cEd124A474f051f9744a301C26D1017B3D54 \\
      & Psychedelics Anonymous Genesis & 0x75E95ba5997Eb235F40eCF8347cDb11F18ff640B \\
      & Art Blocks &  0xa7d8d9ef8D8Ce8992Df33D8b8CF4Aebabd5bD270 \\
      & CoolmansUniverse & 0xa5C0Bd78D1667c13BFB403E2a3336871396713c5 \\
      & My Pet Hooligan & 0x09233d553058c2F42ba751C87816a8E9FaE7Ef10 \\
      & Decentraland & 0xF87E31492Faf9A91B02Ee0dEAAd50d51d56D5d4d \\
      & Terraforms &  0x4E1f41613c9084FdB9E34E11fAE9412427480e56 \\
      & Doodles &  0x8a90CAb2b38dba80c64b7734e58Ee1dB38B8992e \\
      & Neo Tokyo &  0x698FbAACA64944376e2CDC4CAD86eaa91362cF54 \\
      & Creature World &  0xc92cedDfb8dd984A89fb494c376f9A48b999aAFc \\
      & More Than Gamers &  0x49907029e80dE1cBB3A46fD44247BF8BA8B5f12F \\
      & VOX Series 1 &  0xad9Fd7cB4fC7A0fBCE08d64068f60CbDE22Ed34C \\
      & DinoBabies &  0xf9d53E156fE880889E777392585FEb46D8D840f6 \\
      & Party Bears &  0x35471f47c3C0BC5FC75025b97A19ECDDe00F78f8 \\
      & DEGENERATE/REGENERATE &  0x7828c811636CCf051993C1EC3157b0B732e55B23 \\
      & ALIENFRENS &  0x123b30E25973FeCd8354dd5f41Cc45A3065eF88C \\
      & NFT Worlds &  0xBD4455dA5929D5639EE098ABFaa3241e9ae111Af \\
      & CyberKongz &  0x57a204AA1042f6E66DD7730813f4024114d74f37 \\
      & Slotie &  0x5fdB2B0C56Afa73B8ca2228e6aB92Be90325961d \\
      & RumbleKongLeague &  0xEf0182dc0574cd5874494a120750FD222FdB909a \\
      & Cool Cats &  0x1A92f7381B9F03921564a437210bB9396471050C \\
      & Punks Comic 2 &  0x128675d4FddbC4a0D3f8aA777D8EE0fb8B427C2F \\
      & Wolf Game &  0x7F36182DeE28c45dE6072a34D29855BaE76DBe2f \\
      & Little Lemon Friends &  0x0B22fE0a2995C5389AC093400e52471DCa8BB48a \\
      & VeeFriends &  0xa3AEe8BcE55BEeA1951EF834b99f3Ac60d1ABeeB \\
      & Crypto Bull Society &   0x469823c7B84264D1BAfBcD6010e9cdf1cac305a3 \\
      & House of Legends &   0x8C714199d2eA08CC1f1F39A60f5cD02aD260A1e3 \\
      & Meta-Legends &  0xF9c362CDD6EeBa080dd87845E88512AA0A18c615 \\
      & Gambling Apes &   0x90cA8a3eb2574F937F514749ce619fDCCa187d45 \\
      & BoredApeKennelClub &  0xba30E5F9Bb24caa003E9f2f0497Ad287FDF95623 \\
      & Cryptoadz &   0x1CB1A5e65610AEFF2551A50f76a87a7d3fB649C6 \\
      & apekidsclub &   0x9Bf252f97891b907F002F2887EfF9246e3054080 \\
      & ExpansionPunks &   0x0D0167A823C6619D430B1a96aD85B888bcF97C37 \\
      & Elderly Ape Retirement Club &   0x9ee36cD3E78bAdcAF0cBED71c824bD8C5Cb65a8C \\
      & Divine Wolves &   0xb4e9123bd3Ef4Df17f8cc6EF7C2Be66428CF4931 \\
      & Superlative Apes &   0x1e87eE9249Cc647Af9EDEecB73D6b76AF14d8C27 \\
      & Desperate ApeWives &   0xF1268733C6FB05EF6bE9cF23d24436Dcd6E0B35E \\
      & Solarbots &  0x8009250878eD378050eF5D2a48c70E24EB2edE7E \\
      & merge. &   0xc3f8a0F5841aBFf777d3eefA5047e8D413a1C9AB \\
      & Neo Tokyo Part 3 &   0x0938E3F7AC6D7f674FeD551c93f363109bda3AF9 \\
      & Art blocks &   0x059EDD72Cd353dF5106D2B9cC5ab83a52287aC3a \\
      & Champions &   0x97a923ed35351a1382E6bcbB5239fc8d93360085 \\
      & Pepsi Mic Drop &   0xa67D63E68715DCF9b65e45e5118b5fcD1e554b5f \\
      & CROAKZ &  0x7caE7B9b9a235D1D94102598E1f23310A0618914 \\
      & Meebits &  0x7Bd29408f11D2bFC23c34f18275bBf23bB716Bc7 \\
      & Cyber Gorillas &  0x1E1b4E127A510cafa6d0eAec024a4319a5E18821 \\
      & Lost Poets &   0x4b3406a41399c7FD2BA65cbC93697Ad9E7eA61e5 \\
      & SuperRare &   0xb932a70A57673d89f4acfFBE830E8ed7f75Fb9e0 \\
      & Lazy Lions &   0x8943C7bAC1914C9A7ABa750Bf2B6B09Fd21037E0 \\
      & Neo Tokyo Part 2 &  0xab0b0dD7e4EaB0F9e31a539074a03f1C1Be80879 \\
      & Loot &  0xFF9C1b15B16263C61d017ee9F65C50e4AE0113D7 \\
      & Cryptovoxels &   0x79986aF15539de2db9A5086382daEdA917A9CF0C \\
    \hline
    Metaverse & Decentraland & 0xF87E31492Faf9A91B02Ee0dEAAd50d51d56D5d4d \\
    & Sandbox & 0x50f5474724e0Ee42D9a4e711ccFB275809Fd6d4a \\
    & Cryptovoxels & 0x79986aF15539de2db9A5086382daEdA917A9CF0C \\
    & NFT Worlds & 0xBD4455dA5929D5639EE098ABFaa3241e9ae111Af \\
    \end{tabular}
    \caption{Composition of the indices.}
    \label{tab:index_compo}
\end{table}
\end{document}